# Does AI and Human Advice Mitigate Punishment for Selfish Behavior? An Experiment on AI ethics From a Psychological Perspective

Margarita Leib[1], Nils Köbis[2,3], & Ivan Soraperra[2]

1 - *Department of Social Psychology, Tilburg University*

2 – *Center for Humans and Machines, Max Planck Institute for Human Development*

3 – *Research Center Trustworthy Data Science and Security, University Duisburg-Essen*

Forthcoming in *Computers in Human Behavior*

Corresponding Author:

Margarita Leib, Tilburg University, Department of Social Psychology, P.O. Box 90153, 5000 LE Tilburg, The Netherlands. Email: m.leib@tilburguniversity.edu

Author Contributions: All authors conceptualized the project, designed the experiment, and wrote the manuscript. M.L collected and analyzed the data.

Acknowledgement: The study was supported by funding from the research center trustworthy data science and security, university of Duisburg-Essen; the department of social psychology, Tilburg university; and the Center for Humans and Machines, MPI for Human Development. We further wish to thank Dianna Amasino as well as all the participants of the "Machine Behavior Research on Cooperative AI" workshop for their helpful feedback.

## Abstract

People increasingly rely on AI-advice when making decisions. At times, such advice can promote selfish behavior. When individuals abide by selfishness-promoting AI advice, how are they perceived and punished? To study this question, we build on theories from social psychology and combine machine-behavior and behavioral economic approaches. In a pre-registered, financially-incentivized experiment, evaluators could punish real decision-makers who (i) received AI, human, or no advice. The advice (ii) encouraged selfish or prosocial behavior, and decision-makers (iii) behaved selfishly or, in a control condition, behaved prosocially. Evaluators further assigned responsibility to decision-makers and their advisors. Results revealed that (i) prosocial behavior was punished very little, whereas selfish behavior was punished much more. Focusing on selfish behavior, (ii) compared to receiving no advice, selfish behavior was penalized more harshly after prosocial advice and more leniently after selfish advice. Lastly, (iii) whereas selfish decision-makers were seen as more responsible when they followed AI compared to human advice, punishment between the two advice sources did not vary. Overall, behavior and advice content shapes punishment, whereas the advice source does not.

**Does AI and Human Advice Mitigate Punishment for Selfish Behavior? An Experiment on AI Ethics From a Psychological Perspective**

In October 2024, an attorney submitted legal documents containing false information that benefited him at the expense of others. An investigation revealed that the attorney had relied on AI tools for guidance and advice, prompting a judge to recommend a $15,000 fine (Merken, 2025). Similar cases have emerged in recent years, with lawyers fined anywhere from $1,000 to $15,000 for relying on AI-generated advice (Merken, 2025; Cole, 2025). These incidents highlight that people do follow and misuse AI advice for personal gain (see also Leib et al., 2024), that such behavior is punished, and that the severity of punishment varies widely. This raises critical questions: How is responsibility shared between human decision-makers and their AI advisors? Importantly, to what extent are individuals punished for acting on selfish AI advice? The current work tackles the emerging question of punishment in hybrid human-AI settings.

The increasing integration of AI into human decision-making has sparked a growing debate about AI's culpability (Rahwan et al., 2019). Whereas some suggest AI is a moral agent, depending on its features (Maruyama, 2022) or the situation (Heinrichs, 2022), others propose that AI cannot be morally responsible at all (Constantinescu et al., 2022). This debate is no longer purely philosophical, as legal scholars also stress the emerging need to adapt current legal frameworks to account for the role and liability of AI in hybrid human-AI decisions (Giuffrida, 2019; Soyer & Tettenborn, 2022).

Current legal frameworks recognize the crucial influence of human advisors when assigning responsibility and determining punishment. For example, individuals who were coerced into wrongdoings by humans are often treated as less culpable, resulting in lower criminal liability (Barnett & Becker, 1987; Dressler, 2001; Feinberg, 1989). However, whether individuals coerced by AI will receive a similar treatment remains unclear. Because laws around liability and punishment in human-AI settings are still developing, many legal claims featuring AI are being decided on a case-by-case basis (Smith et al., 2024). As such, examining how the general public assigns responsibility and punishes selfish behavior in hybrid human-AI settings bears immense relevance. Understanding public preferences is essential for developing informed and widely acceptable legal frameworks, as the general public can influence legal decisions (e.g., as jurors in the USA) and plays a crucial role in supporting new laws and regulations.

To provide initial insights into this topic, we examine: (i) the extent to which selfish behavior after receiving AI advice is punished; (ii) how this punishment compares to the punishment of selfish behavior after receiving human advice or no advice; and as a control condition, (iii) how these punishment patterns compare to those for prosocial behavior after receiving AI, human, or no advice.

**AI ethics: A psychological approach**

Discussions on the ethics of technology started as early as the invention of computers (Brey & Søraker, 2009). With the introduction of AI, the debate on AI ethics developed, initially focusing on technical aspects, addressing questions around how AI should be developed and used in order to uphold moral principles (Mittelstadt et al., 2016; Russell, 2019). This classical approach was normative, prescribing normative standards around how AI *should* be programmed -- for instance aligned with human values (Christian, 2021) – when affecting people's ethics, rather than how AI *actually* shapes it (Jobin et al. 2019; Wallach & Allen, 2008). For example, Jobin and colleagues (2019) analyzed reports and guidance documents for ethical AI around the world in order to understand what governments and organizations deem important for the realization of ethical AI. Results revealed five key principles: transparency, justice and fairness, non-maleficence, responsibility, and privacy. Furthermore, the authors found that although these principles were deemed important across the globe, the exact meaning and interpretation of each principle and what should be done to uphold it differed substantially.

This normative approach for AI ethics helps to set goals and guidelines for how AI should function while preserving moral principles. It further helps to identify where AI systems go wrong, and what should be done to fix them. For instance, research found that although promising fairness by removing subjective human biases, AI systems often inadvertently reinforce existing biases. Algorithmic decisions based on biased datasets can exacerbate discrimination, racism, and gender inequality (Kleinberg et al., 2018; Obermeyer et al., 2019). The findings emphasize the importance of carefully selecting training data and fine-tuning AI models to mitigate these risks.

Recently, work on AI ethics shifted from a purely normative approach to a descriptive one, examining how AI *actually* influences ethical decision-making, rather than how it ought to. A similar transition occurred in the study of ethics more generally: whereas early research in ethics was normative, prescribing how people should behave when facing a temptation to be unethical (Trevino & Weaver, 1994), the last fifteen years have seen a shift towards a descriptive approach, focusing on how individuals actually behave when faced with ethical dilemmas (Bazerman & Gino, 2012; Shalvi et al., 2015).

Focusing on the human aspect in human-AI interactions, this research explores how individuals perceive and interpret AI, as well as how these interactions shape ethical judgment and behavior (Awad et al. 2018; Awad et al. 2020; Leib et al., 2024). Given the focus on humans' experience, this work draws on foundational insights from psychological research (Köbis et al. 2021; Bonnefon et al. 2024; Irlenbusch et al. 2025). Integrating psychology into AI ethics provides insight into the cognitive, emotional, and social factors that shape humans' engagement with AI. This in turn allows a deeper understanding of how AI influences human ethical decision-making, and the underlying psychological mechanisms driving these effects.

Key research focusing on humans in AI ethics assesses attitudes towards AI in ethical dilemmas. Examining preferences across a variety of contexts (e.g., legal, medical) Bigman and Gray (2018) revealed that individuals are rather averse to machines making moral decisions and prefer humans to be responsible when decisions entail ethical considerations. Other seminal work examined preferences for moral decisions made by autonomous vehicles across the globe. The findings revealed large variations in moral preferences that map onto distinct cultural clusters, highlighting how cultural and societal contexts shape ethical judgments (Awad et al. 2018).

Focusing on unethical behavior, recent work demonstrated that the ability to delegate to AI systems increases dishonesty. Namely, when faced with a temptation to lie for financial profit, individuals engage in more dishonesty when delegating such decisions to AI than when they decide for themselves (Köbis et al. 2024). Moreover, the programming interface used to instruct AI delegates substantially influences dishonesty levels. Interfaces based on goal specification or selection of training data resulted in higher levels of dishonesty compared to interfaces programmed with explicit rules. The reason is that selection of a goal or data introduces greater ambiguity to the decision-making process, serving as a justification for ethical misconduct (Pittarello et al., 2015; Leib et al., 2019).

Whereas previous research has examined how AI influences ethical decision-making, such as increasing dishonesty via delegation (Köbis et al., 2024) or shaping moral judgments (Bigman & Gray, 2018; Awad et al., 2018), little is known about its role in punishment decisions. Following a descriptive approach to AI ethics, we examine how adherence to AI advice, particularly when it promotes selfish behavior, affects punishment decisions.

## The Growing Influence of AI Advice on Human Behaviour

AI-enabled technologies have taken various roles in society (Rahwan et al., 2019), leading to new ethical challenges (Floridi & Sanders, 2004; Köbis et al., 2021; Starke et al., 2024). As an advisor, AI can influence consumer preferences (Werner et al., 2024; Fink et al., 2024) and sway financial decisions (Klingbeil et al., 2024). Moreover, AI advice affects people's moral perceptions and behavior. For instance, participants' judgments of whether to sacrifice one life to save five, a classic moral dilemma, is influenced by ChatGPT-generated advice: when advised that one person should be sacrificed, the majority of participants report they believe that as well. When advised that one person should not be sacrificed, only a minority disagree (Krügel et al., 2023). Similarly, after reading AI advice that promotes dishonest behavior, participants lie more to secure a higher financial gain (Leib et al., 2024; see systematic literature review, Poszler & Lange, 2024).

While the influence of AI on human decision-making has been extensively documented, the punishment of individuals who follow bad AI advice remains largely unexplored. To address this gap, our study combines approaches and insights from multiple disciplines: behavioral economics, machine behavior, and social psychology. First, we draw on behavioral economics by using a financially incentivized, behavioral measure of punishment. By doing so, we supplement previous work that focused on stated preferences (Lima et al., 2021; Malle et al., 2025; Awad et al., 2020; Wilson et al., 2022) with a measure that captures costly, behavioral punishment (Fehr & Fischbacher, 2004). Costly punishment serves as a useful behavioral proxy because it reflects realistic situations in which punishment has tangible consequences for both the punisher and the target. Compared to self-report measures, it has greater internal validity (Roe & Just, 2009) and is less vulnerable to social desirability bias (van de Mortel, 2008).

Second, we adopt a machine behavior approach (Rahwan et al., 2019). We use real, state-of-the-art algorithms to generate AI advice and examine punishment reactions to behaviors that stem from AI advice in the same way we study punishment reactions to behaviors that stem from human advice (Rahwan et al., 2022). This approach allows us to avoid unnecessary experimental deception (Rahwan et al., 2022), as well as to study AI in the same way we study humans, a key aspect of the machine behavior approach (Rahwan et al., 2019). Third, we draw on work from social psychology. Namely, we build on a key theoretical framework from social psychology – attribution theory (Heider, 2013) – to formulate our hypotheses. When building our hypotheses, we further supplement the theory with recent work examining hypothetical responses to AI and human wrongdoings.

In the subsequent sections, we begin with a general introduction to punishment, followed by our pre-registered hypotheses. In line with the order of our pre-registration, we first present hypotheses about the punishment of selfish versus prosocial behaviors. Next, we move to hypotheses regarding the impact of receiving advice, in general, on punishment. Finally, we flesh out hypotheses regarding the effects of receiving AI versus human advice on punishment.

**Punishment and Moral Judgment of Humans and Machines**

The willingness to punish others for their transgressions is an integral part of our moral sensitivities (Hofmann et al., 2018). The idea that one's wrongdoings may be punished helps to deter individuals from engaging in such wrongdoings and promotes behavior aligned with positive social norms. Indeed, punishment is prevalent across societies and cultures (Henrich et al., 2006; Balliet & Van Lange, 2013) and helps signal which behaviors are acceptable and which are not. Simply put, punishment ensures that individuals adhere to social expectations and norms (Gachter et al., 2008; Fehr & Fischbacher, 2004).

Punishment can be found in various fields of psychology. Starting from behaviorism, Skinner's Reinforcement theory conceptualized punishment as an external consequence that reduces the occurrence of an undesired behavior (Skinner, 1965). Over time, the concept of punishment expanded beyond its behaviorist roots and took on more complex and sometimes subtle forms. In cognitive psychology, for instance, negative feedback can be seen as a form of subtle punishment, signaling disapproval of one's actions (Kluger & DeNisi, 1996). Similarly, in the constructivist approaches, negative feedback (or the absence of positive one) helps individuals actively learn and construct their understanding of what are socially acceptable behaviors (Vygotsky, 2011).

More recently, Molho and colleagues (2020) examined punishment strategies in daily life using a diary study approach. Their findings revealed that in response to norm violations, participants engage in direct confrontation, gossip, and avoidance of the offender. A key insight from this work is that when punishing, individuals engage in a cost-benefit analysis. For instance, participants were more likely to directly confront offenders that personally hurt them, but less likely to confront those in higher power positions (Molho et al., 2020). Building on the idea that punishment involves a cost-benefit analysis, we focus on a behavioral measure of *costly punishment* (Fehr & Fischbacher, 2004), where individuals need to sacrifice part of their financial endowment to punish others.

Ample research in behavioral science consistently demonstrates that selfish actions are judged more harshly and are punished more than prosocial actions. Indeed, when individuals do not cooperate, offer unfair monetary splits to others, or behave antisocially, they are likely to be punished (Fehr & Fischbacher, 2004; Pillutla & Murnighan, 1996). As punishment serves as a deterrent and helps to enforce positive social norms, the tendency to punish antisocial behavior stems from societal norms favoring cooperation, fairness, and prosociality as foundations for social cohesion (Fehr & Fischbacher, 2004; Balliet et al., 2011; Balliet & Van Lange, 2013; Clutton-Brock & Parker, 1995). Thus, we hypothesize that:

**(H1)** selfish behavior will be punished more than prosocial behavior,

**(H2a)** prosocial behavior will virtually not be punished, and subsequently:

**(H2b)** the type of advice received prior to prosocial behavior will not affect punishment.

Focusing on selfish behavior, punishment may depend on how people perceive the person acting selfishly and the possible reasons behind selfishness. Attribution theory explains how individuals interpret the causes of others' behavior, particularly in moral and social contexts. According to the theory, people tend to attribute actions either to internal dispositions, such as personality traits or beliefs, or to external factors, such as situational or environmental features (Heider, 2013; Kelley, 1973). When considering to what extent to punish someone for their (bad) actions, individuals take into account the perceived intentionality and autonomy behind the act. Behaviors seen as freely chosen and intentional tend to be judged more harshly, whereas those perceived as externally influenced tend to be excused or punished less severely (Weiner, 1996).

Applying attribution theory to the context of punishing selfish behavior, the type of advice decision-makers receive before acting selfishly can influence attributions of intent and, consequently, the severity of punishment. As previous research suggests, harmful actions are judged not only based on their outcomes but also based on the perceived intentionality of the decision-maker (Cushman, 2008; Heider, 2013; Kelley, 1973). Perceptions of autonomy and intention are shaped by coercion (Monroe et al., 2015) and persuasion (Pavey & Sparks, 2009). Decisions made without any coercion and persuasion are perceived as more intentional than those that occurred after persuasion (Monroe et al., 2015; Pavey & Sparks, 2009). As such, compared to selfish behavior that occurred in the absence of advice, selfish behavior that occurred after advice promoting selfishness would be seen as less intentional and might be punished less.

In contrast, if external pressures such as coercion and persuasion fail, behavior may be perceived as even more likely to stem from internal disposition (Heider, 2013; Kelley, 1973). Therefore, selfish behavior that occurs after advice promoting prosociality will be perceived as even more intentional and might be punished even more. Indeed, ignoring prosocial advice and engaging in selfish behavior means deliberately rejecting a moral norm and expectation, which in turn increases condemnation (Cushman, 2008). Taken together, we hypothesize that:

**(H3a)** compared to receiving no advice, selfish behavior will be punished less after receiving selfish advice, and

**(H3b)** compared to receiving no advice, selfish behavior will be punished more after receiving prosocial advice[1].

Beyond the content of advice, the source of the advice received – whether from an AI or a fellow human – might affect the punishment of selfish behavior. Research on how people react to AI advice, especially when it appears in a human-like text, has only recently emerged. As such, work on the topic is in its early stages, and results are mixed.

---

[1] In the pre-registration, we further outlined the same hypotheses focusing on human-advice. Namely, compared to no advice, selfish behavior will be punished less after receiving a selfish human-advice; and punished more after receiving prosocial human-advice. To increase readability and succinctness we focus on the hypotheses comparing AI and human advice in the rest of the introduction and report analyses for all pre-registered hypotheses in the Results section.

On the one hand, a theoretical machine ethics account argues that machines not only assist humans in making moral decisions but also carry moral responsibility by themselves (Van de Voort et al., 2015). In line with this view, recent studies reveal that when AI agents or humans engage in wrongdoing or moral violations, both are perceived as responsible, sometimes to similar extents (Lima et al., 2021; Malle et al., 2025). Furthermore, when making ethical decisions, people follow AI and human advice to a similar extent (Carrasco-Farre, 2024; Leib et al., 2024) and feel the same level of shared responsibility when advised by humans or AI (Leib et al., 2024). This initial body of evidence might suggest that punishment will not differ between the two advice sources.

However, competing evidence suggests that punishment may differ between the human and AI advice sources. First, in line with attribution theory (Heider, 2013; Kelley, 1973; Weiner, 1996), the more intentional a behavior seems, the more moral responsibility one is perceived as having for it. Compared to humans, AI is often viewed as lacking independent intent or moral reasoning and, as such, as less culpable. Indeed, when evaluating mistakes with moral consequences (Awad et al., 2020) or immoral decisions (Wilson et al., 2022), humans are perceived as more blameworthy than machines. Similarly, when humans discriminate, people are more morally outraged than when algorithms do the same (Bigman et al., 2023).

Punishment following AI advice

Because of the higher perception of intent to humans compared to machines, people may attribute more responsibility to a human advisor than an AI advisor when deciding how much to punish selfish behavior. If that is the case, a higher share of the responsibility will be shifted *away* from the selfish decision-maker (and towards the advisor) when the advisor is human (vs. AI), resulting in less punishment. Recent work found that people intuitively apply a similar logic when choosing an advisor. Specifically, in organizational settings, when motivated to share responsibility for their decisions, experienced managers prefer receiving advice from humans instead of algorithms (Aschauer et al., 2023). Taken together, we hypothesize that:

**(H4a)** selfish behavior will be punished less after receiving selfish human advice compared to selfish AI advice, and

**(H4b)** selfish behavior will be punished more after receiving prosocial human advice compared to prosocial AI advice.

## Method

### Overview of the Experimental Stages

The experiment entailed three stages (see Figure 1). In the *Advice Writing* stage, we collected advice texts written by humans and AI. In the *Decision-Making* stage*,* we collected selfish and prosocial behavior by participants after they read an advice text (written by a human or an AI) or after they did not read any advice. In the *Evaluation* stage*,* the main part of our study, we collected the punishment decisions for the behaviors we collected in the *Decision-Making* stage. Below, we elaborate on each of the stages.

### Part 1 - Advice Writing stage

**Human advice.** To collect human advice, we recruited a sample of 51 participants ($M_{age}$ = 28.74, $SD_{age}$ = 8.06, 45.09% females) via Prolific.co to serve as advisors. The task took about 7 minutes, and participants received £1 for completing it. Advisors learned that a separate group of participants (decision-makers) would later engage in an investment task in which they would choose between a selfish option (A) and a prosocial option (B). Advisors were instructed to write two advice texts for these decision-makers, one encouraging decision-makers to choose the selfish option (A) and another encouraging decision-makers to choose the prosocial option (B). The order in which advisors were asked to write the texts was randomized.

To incentivize advisors to write convincing texts, they were informed that among all advisors, 10% would be randomly selected and a decision-maker would read their advice. If a decision-maker who read their advice followed it, they would earn an additional bonus of £2. That is, (i) if a decision-maker reads their selfish advice and chooses the selfish option or (ii) if a decision-maker reads their prosocial advice and chooses the prosocial option, they would earn £2. Otherwise, they would earn £0.

**AI advice.** To collect AI-generated advice texts, we provided instructions to GPT-4 (version:0613) comparable to those given to human advisors. After collecting all human and AI advice, we screened the advice texts to make sure they were in line with the provided instructions. We then randomly selected five advice texts to implement in the decision-making stage. For more details about the screening process, see the supplementary online materials (SOM). For the full instructions, the prompts used for GPT-4, and all human and AI advice texts, see Open Science Framework (OSF).

***Part 2 - Decision-Making stage***

In the *Decision-Making* stage*,* we collected a separate sample of 60 participants ($M_{age}$ = 40.30, $SD_{age}$ = 12.56, 68.33% females) via Prolific.co (pre-registration) to act as decision-makers. The task took about 6 minutes, and decision-makers earned £1 for participating in the study. Decision-makers engaged in an investment task, which affected the distribution of points (later converted to money) between themselves and a charity (UN carbon offset platform, used to offset carbon emissions). Specifically, decision-makers could choose between options A and B. Option A was selfish, giving decision-makers 100 points and the charity 0 points. In contrast, option B was prosocial, giving decision-makers 60 points and the charity 40 points[2]. At the end of the study, the points were converted to money, such that 100 points = £4, and paid out to all decision-makers and the charity.

Before deciding, decision-makers either read an advice text written by an AI or a human. The advice was either selfish, suggesting to decision-makers that they should choose option A, or prosocial, suggesting to them to choose option B. In a control condition, decision-makers did not read advice before making their investment choice. Overall, for the decision-makers in the *Decision-Making* stage, we implemented a 2 (AI advice vs. human advice) by 2 (selfish advice vs. prosocial advice) + 1 (no advice) between-subjects design. Note that decision-makers' behavior is not the focus of our investigation and that their actions were collected only to implement punishment choices in the next stage[3]. For the complete instructions for decision-makers, see OSF.

---

[2]Note that the labels 'selfish' and 'prosocial' are relative. That is, option A is more selfish than option B, and option B is more prosocial than option A. In a follow-up survey, we asked 200 participants from the original sample of evaluators (participants who took part in part 3 – the evaluation stage) about their perceptions of options A and B on a scale (0 = selfish to 100 = prosocial). Indeed, participants reported lower values for option A ($M$ = 20.52, $SD$ = 22.91) than for option B ($M$ = 76.72, $SD$ = 21.35), paired t-test: $t(199)$ = -21.29, $p$ < .001, Cohen's $d$ = 2.539, indicating that they perceive option A as more selfish (and less prosocial) than option B. See SOM for details.

### *Part 3 - The Evaluation stage*

The *Evaluation* stage is the main focus of our paper. In it, another group of participants (henceforth evaluators) read about the decision-makers' choices and could punish them. Evaluators first read about the investment task that decision-makers completed. Then, evaluators learned that they would observe and evaluate different behaviors by different decision-makers. Specifically, evaluators learned about (i) the type of advice the decision-maker read (no advice, prosocial advice, or selfish advice), (ii) the source of advice (no advice, AI, or human), and (iii) the decision-makers' behavior in the investment task (prosocial or selfish). Then, for every decision-maker they observed, evaluators engaged in the costly punishment task.

**Information about advice type.** When punishing, evaluators learned about the type of advice that decision-makers received. Specifically, they learned whether the advice was selfish (suggested choosing option A) or prosocial (suggested choosing option B). Evaluators did not read the content of the (AI or human) advice. We chose not to disclose the exact content of the advice to isolate the effect of the advice's source (AI or human) and type (selfish or prosocial) on punishment. This approach ensures that punishment decisions are based entirely on evaluators' judgment of how selfish (and prosocial) decision-makers should be treated, abstracting away from the possible content of AI and human advice. See the full instructions on OSF.

---

[3]When making their investment, decision-makers knew that another group of participants (evaluators) would observe their choice and could reduce their points by a maximum of 60 points. Decision-makers knew that when making their choices, evaluators would learn about the source of the advice (no advice, AI, or human), the advice given (no advice, prosocial, or selfish), and decision-makers' behavior (prosocial or selfish).

**Information about the advice source.** When punishing, evaluators learned whether decision-makers received advice and, if so, whether a human or an AI had written the advice. Evaluators received information about the way the advice was collected. Specifically, when they evaluated decision-makers who received human-written advice, they read:

> "A separate group of participants wrote advice regarding the investment decision. These participants received a set of instructions to help them craft their advice (the same instructions were given to the AI as well). They did not participate in the investment task themselves, only wrote advice for decision-makers."

Similarly, when they evaluated decision-makers who received AI-written advice, they read:

> "The algorithm was trained on large datasets of English text (> 820 GiB, more than 500 million pages of text) and produces text by predicting the next word in a sentence (similar to predictive text on smartphones). The AI was given the same instructions as the human advisors to generate its advice."

**Full experimental design.** Overall, evaluators observed all possible combinations of behavior and advice: 5 (advice options: [human/AI × selfish/prosocial advice + no advice]) by 2 (decision-makers' behavior: [selfish/prosocial]), and made 10 punishment decisions - one for every combination. The order was randomized such that the source of advice (AI vs. human vs. no advice) and type (prosocial vs. selfish advice) were randomized between blocks, and decision-makers' behavior (prosocial vs. selfish) was randomized within a block. Note that the main focus of this work is the punishment of selfish behaviors, and that the punishment of prosocial acts serves mainly as additional control conditions.

## Measures

***Costly punishment.*** To capture costly punishment, we implemented a commonly used costly punishment paradigm (Fehr & Fischbacher, 2004). In the task, evaluators received 70 points. They could spend up to 20 points to deduct points from the decision-maker's final payoff. For every point an evaluator spent, the corresponding decision-maker lost 3 points. The evaluators' actions did not affect the charity donation. At the end of the study, the points were converted to additional bonus pay (100 points = £4) and paid out to the evaluators and the corresponding decision-makers.

For instance, if a decision-maker choose option A (100 points to self, 0 to UN carbon offset platform), and the evaluator spent 10 of their points to punish the decision-maker, the final payoffs were: Decision-maker: 100 - (10 * 3) = 70 points = £2.8; UN carbon offset platform: £0; Evaluator: 70 − 10 = 60 points = £2.4.

The strength of employing the costly punishment paradigm (Fehr & Fischbacher, 2004) is that it is a behavioral task that captures real and costly behavior. While the task may not capture all the diverse ways that individuals punish one another, this measure is *costly to the punisher* and *consequential to the one being punished*. That is, the evaluator has to be willing to spend some of their own money to punish. This money reflects other costs people are willing to pay in other real-life settings to punish, such as effort, time, or social costs. Furthermore, the decision to punish has real consequences for the one being punished. The more an evaluator punishes, the more money the decision-maker will lose. The loss of money from the decision-maker's side reflects other costs people may incur when punished in real life, such as financial opportunities or good reputation.

***Perception of responsibility.*** After each punishment decision, evaluators further reported the extent to which they thought the decision-maker and advisor were responsible. Specifically, evaluators reported, on a scale from 0 to 100, how they attributed responsibility between the decision-maker and the advisor (0 = the decision-maker is fully responsible; 50 = the decision-maker and advisor share responsibility equally; 100 = the advisor is fully responsible). This measure, and the reported results, are exploratory.

*Additional measures.* For every decision-maker, evaluators further indicated on a scale from 0 (=not at all) to 100 (=very much so), the extent to which they perceived the decision-maker's behavior to be (i) justifiable; (ii) socially appropriate, and (iii) common. Finally, at the end of the task, participants reported on a scale from 0 (=not at all) to 100 (=very much so) the extent to which they (i) are familiar with and (ii) use AI-based large language models such as ChatGPT. These measures, and the results (reported in the SOM), are also exploratory.

**Comprehension and 'reminder' questions.** After reading the instructions, evaluators answered three comprehension questions and another attention check. If they answered a comprehension question incorrectly, they could try again. If they answered the same question incorrectly twice, or if they answered the attention check incorrectly, they were disqualified from participating in the study.

Further, before every punishment decision, evaluators answered three questions to ensure that they read and remembered the relevant information about the decision-maker they were about to punish. These 'reminder' questions asked: (i) what did the advice encourage? (option A, B, or no advice), (ii) what was the source of advice? (Human, AI, or no advice), and (iii) what did the decision-maker choose? (option A or B). Participants who answered at least one of these questions wrong were reminded about the type and source of the advice, as well as the decision-maker's choice. Only after the additional reminder could evaluators make their punishment decisions by selecting how many points (between 0 and 20) they wished to spend to deduct points from the decision-maker. Importantly, and in line with our pre-registration, these additional

'reminder' questions were not used to exclude participants but simply to remind them of key details about the decision-maker they were evaluating.

**Sample and power calculations.** The evaluators' task took about 18 minutes to complete, and participants earned a fixed pay of £2.70 for participating in the study. We pre-registered to collect a sample of 600 participants via Prolific.co. The sample size was determined based on a power analysis to ensure adequate power for capturing the difference in punishment between selfish behavior after receiving AI vs. human advice (H4a and H4b) against the null hypothesis of no such difference. Specifically, when comparing AI advice to human advice, the sample size allows us to achieve 91% power to simultaneously detect (i) a reduction of one punishment point (out of 20) after receiving prosocial advice and (ii) an increase of one punishment point after receiving selfish advice.

Overall, we collected a representative USA sample of 633 participants ($M_{age}$ = 45.09; $SD_{age}$ = 15.94, 49.92% females) for the study. These participants passed the comprehension and attention checks, completed the punishment task, and reported their perceptions.

**Payoff implementation.** Punishment was implemented using the strategy method. Evaluators knew that at the end of the study, 10% would be randomly selected to be matched with a decision-maker and paid out. For those who were selected, one punishment choice (corresponding to the specific advice and behavior combination of a

Punishment following AI advice

matched decision-maker) was implemented, affecting the evaluator and the matched decision-maker's bonus pay.

**Ethics and Data Availability Statement**

The IRB board of our department approved the experiment, and all materials, pre-registrations, and data are available on the OSF.

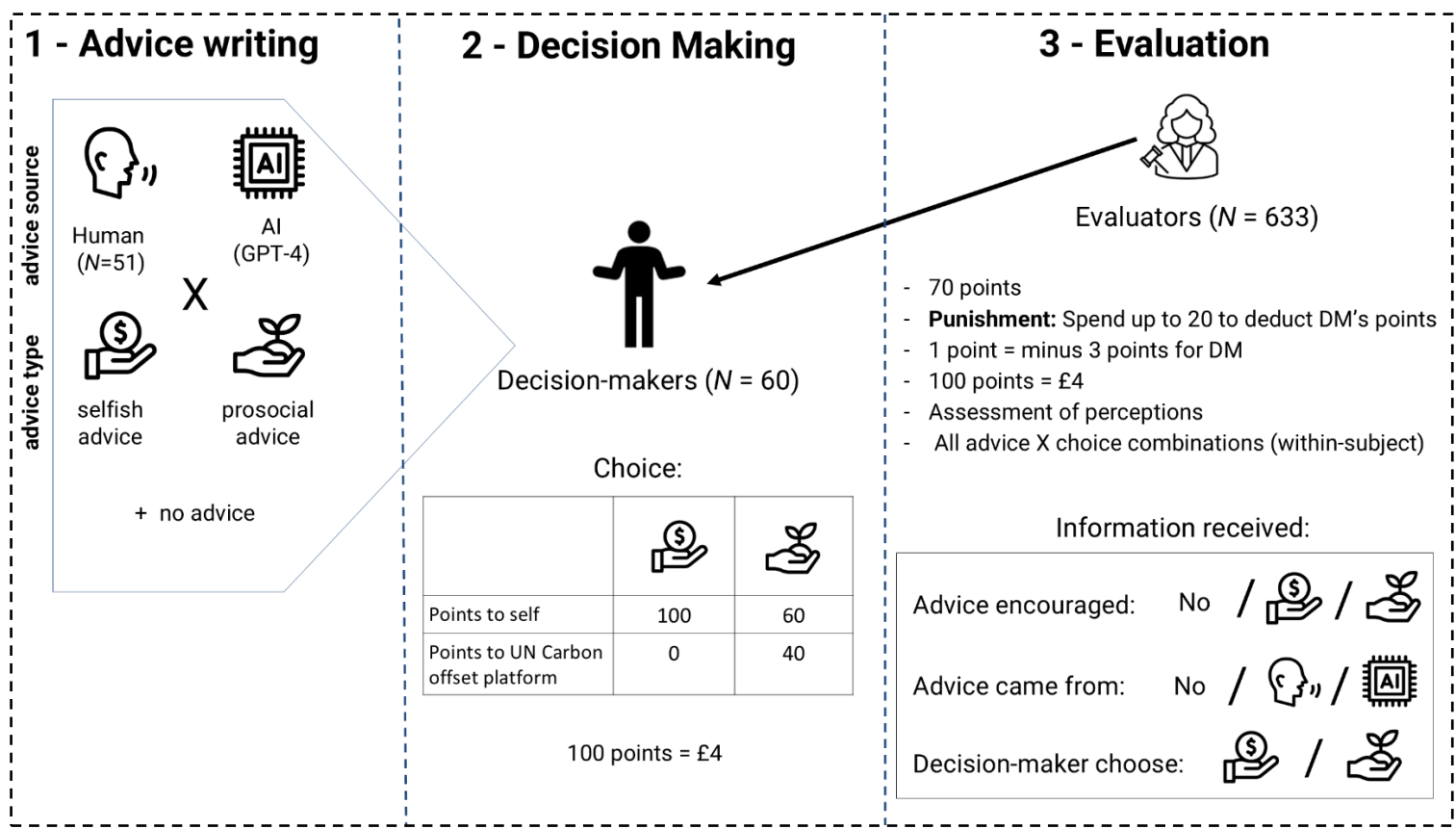

**Figure 1. Overview of experimental setup.** The paper is focused on stage 3 - evaluators' punishment decisions and perceptions.

**Results**

Overall, the points spent to punish decision-makers ranged from 0 to 20, indicating that (some) participants used the entire range of points to punish. On average, evaluators spent 3.62 ($SD$ = 5.35) of their own points, deducting on average 10.86 points (3.62 × 3) from decision-makers.

*(1) Is selfish behavior punished more than prosocial behavior?*

Yes. In line with H1, a linear regression with a random intercept for each participant[4] revealed that evaluators spent more points to punish selfish ($M$ = 5.27, $SD$ = 5.90) than prosocial behavior ($M$ = 1.96, $SD$ = 4.11, $b_{\text{selfish behavior}}$ = 3.304, $p$ < .001, 95% CI = [3.130, 3.478], $\eta_p^2$ = .195), see Figure 2 and Table 1 (model 1).

*(2) Is prosocial behavior punished, and does the type and source of advice matter?*

To some extent, yes. We hypothesized that (H2a) prosocial behavior will virtually not be punished and that (H2b) the type of advice received before prosocial behavior will not affect punishment. However, results reveal that participants punish prosocial behavior, to a small extent. Further, the type of advice affects punishment, yet the advice source does not. Focusing on the punishment of *prosocial* behavior, compared to prosocial decision-makers who received no advice ($M$ = 1.92, $SD$ = 4.06), those who received human-written advice promoting selfish behavior were punished more ($M$ = 2.24, $SD$ = 4.41, $b_{\text{selfish advice}}$ = .321; $p$ = .004, 95% CI = [.100, .540], $\eta_p^2$ = .003).

---

[4]Unless specified otherwise, all reported regressions analyses include a random intercept for each participant.

Further, compared to prosocial decision-makers who received no advice, those who received human-written advice promoting prosocial behavior were punished slightly less ($M$ = 1.17, $SD$ = 3.94, $b_{\text{prosocial advice}}$ = -.207; $p$ = .065, 95% CI = [-.427, .013], $\eta_p^2$ = .001). AI-generated (compared to human-written) advice did not affect punishment decisions. This was the case when decision-makers received selfish advice ($M_{\text{selfish, AI advice}}$ = 2.21, $SD_{\text{selfish, AI advice}}$ = 4.18; $b_{\text{selfish} \times \text{AI advice}}$ = -.025; $p$ = .822; 95% CI = [-.245, .194], $\eta_p^2$ < .001), and prosocial advice ($M_{\text{prosocial, AI advice}}$ = 1.74, $SD_{\text{prosocial, AI advice}}$ = 3.93, $b_{\text{prosocial} \times \text{AI advice}}$ = .030; $p$ = .789; 95% CI = [-.190, .250], $\eta_p^2$ < .001), see Figure 2 and Table 1 (model 2). Thus, whereas participants punished prosocial behavior to a relatively small extent, they punished more when decision-makers engaged in prosocial behavior after receiving selfish advice and less after receiving prosocial advice.

*(3) Is selfish behavior punished, and does the type of advice matter?*

Yes. We hypothesized that compared to receiving no advice, selfish behavior would be punished (H3a) less after receiving selfish advice and (H3b) more after receiving prosocial advice. In line with these hypotheses, we focus on *selfish* behavior and examine how punishment varies based on advice type, pooling the AI and human advice sources together.

Results reveal that compared to selfish decision-makers who received no advice ($M$ = 5.41, $SD$ = 5.97), those who received advice promoting selfish behavior were punished less ($M$ = 4.34, $SD$ = 5.55, $b_{\text{selfish advice}}$ = -1.071; $p$ < .001, 95% CI = [-1.346, -.796], $\eta_p^2$ = .022), and those who received advice promoting prosocial behavior were punished more ($M$ = 6.12, $SD$ = 6.07, $b_{\text{prosocial advice}}$ = .707; $p$ < .001, 95% CI = [.432, .982], $\eta_p^2$ = .010), see Table 1 (model 3). Thus, people punish decision-makers' selfish behavior even more when it occurs after receiving prosocial advice and less when it occurs after receiving selfish advice.

*(4) Does punishment of selfish behavior differ between AI and human advice?*

No. Our final set of hypotheses focused on changes in punishment when advice stems from AI versus humans. Specifically, we hypothesized that (H4a) selfish behavior will be punished less after receiving selfish human advice than selfish AI advice, and (H4b) selfish behavior will be punished more after receiving prosocial human advice than prosocial AI advice. Focusing on the punishment of *selfish* behavior, there was no difference between the punishment of selfish decision-makers who received AI or human advice.

Compared to selfish decision-makers who received no advice ($M$ = 5.41, $SD$ = 5.97), those who received human-written advice promoting selfish behavior were punished less ($M$ = 4.35, $SD$ = 5.52, $b_{\text{selfish advice}}$ = -1.062; $p$ < .001, 95% CI = [-1.378, -.744], $\eta_p^2$ = .017). Further, those who received human-written advice promoting prosocial behavior were punished more ($M$ = 6.21, $SD$ = 6.15, $b_{\text{prosocial advice}}$ = .795; $p$ < .001, 95% CI = [.477, 1.111], $\eta_p^2$ = .009). AI-generated (compared to human-written) advice did not affect punishment decisions. This was the case when decision-makers received selfish advice ($M_{\text{selfish, AI advice}}$ = 4.33, $SD_{\text{selfish, AI advice}}$ = 5.59; $b_{\text{selfish × AI advice}}$ = -.019; $p$ = .907, 95% CI = [-.336, .298], $\eta_p^2$ < .001), and prosocial advice ($M_{\text{prosocial, AI advice}}$ = 6.03, $SD_{\text{prosocial, AI advice}}$ = 6.01, $b_{\text{prosocial × AI advice}}$ = -.175; $p$ = .279; 95% CI = [-.492, .141], $\eta_p^2$ < .001), see Figure 2 and Table 1 (model 4)[5].

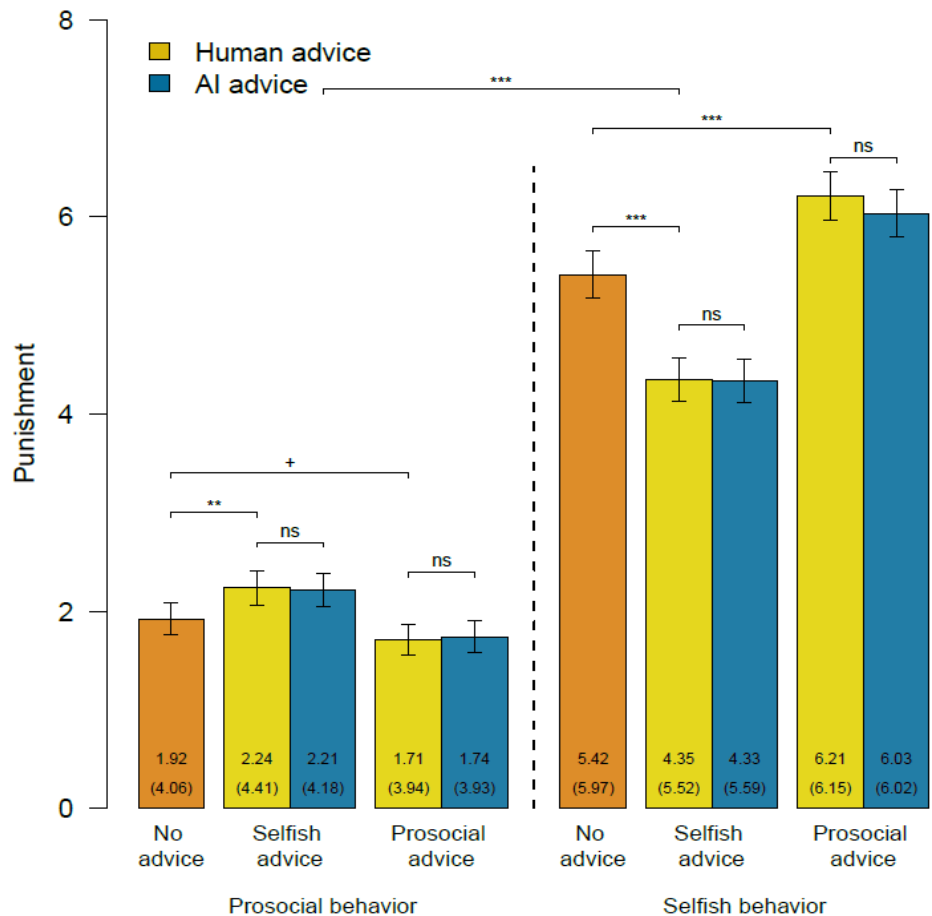

**Figure 2.** Mean number of points spent to punish decision-makers based on their behavior (prosocial or selfish), the advice they received (no advice, prosocial, or selfish), and the source of advice (AI or human). Mean (*SD*) appears at the bottom of each bar; ***$p$ < .001; **$p$ < .01, *$p$ < .05, +$p$ < .1, ns: $p$ >.1.

---

[5]Punishment results are robust to the order in which advice type (selfish or prosocial) and source (AI or human) were presented.

| | Dependent Variable: Points spend on punishment | | | |
|---|---|---|---|---|
| | (1) | (2) | (3) | (4) |
| Selfish behavior | 3.304*** (.089) | | | |
| Selfish advice | | .321** (.112) | -1.071*** (.140) | -1.062*** (.162) |
| Prosocial advice | | -.207+ (.112) | .707*** (.140) | .795*** (.162) |
| Selfish advice × AI advice | | -.025 (.112) | | -.019 (.162) |
| Prosocial advice × AI advice | | .030 (.112) | | -.175 (.162) |
| Intercept | 1.96*** | 1.91*** | 5.41*** | 5.41*** |
| $R^2$ (conditional) | .562 | .765 | .763 | .763 |
| $R^2$ (marginal) | .095 | .003 | .018 | .018 |
| N observations | 6330 | 3165 | 3165 | 3165 |
| N participants | 633 | 633 | 633 | 633 |
| Data used for analysis | All data | Prosocial behavior conditions | Selfish behavior conditions | Selfish behavior conditions |

**Table 1.** Regression analyses on the points spent on punishment (models 1-4). In model 3, the data for AI and human advice are pooled together, making the 'selfish advice' and 'prosocial advice' coefficients reflect the average difference (across advice sources) between 'no advice' and these conditions. In models 2 and 4, the benchmark is 'no advice' and 'selfish advice' and 'prosocial advice' coefficients reflect the gaps between no advice and selfish human advice conditions, and no advice and prosocial human advice conditions, respectively. +$p$ < .10, *$p$ < .05, **$p$ < .01, ***$p$ < .001.

**Reward (and penalty) for advice compliance (and defiance).** So far, our results suggest that both selfish and prosocial decision-makers are punished more when they do not follow the advice they receive and punished less when they follow the advice they receive. We conducted additional analyses to test whether the punishment patterns obtained for selfish decision-makers were solely driven by the fact that decision-makers comply (vs. defy) advice or, alternatively, whether the punishment of selfish decision-makers goes above and beyond simply rewarding advice compliance and penalizing advice defiance.

First, we coded decision-makers' compliance. If decision-makers acted prosocially after receiving prosocial advice or selfishly after receiving selfish advice, they were classified as *compliant*. They were classified as defiant if they acted prosocially after selfish advice or selfishly after prosocial advice. Then, we focused only on the conditions in which advice was present and predicted punishment from compliance (vs. defiance), the type of behavior being punished (prosocial vs. selfish), and the interaction between the two factors.

Results revealed that for prosocial behavior, compliant decision-makers were punished less than defiant decision-makers ($b_{\text{comply with advice}}$ = -.500; $p$ < .001, 95% CI = [-.773, -.227], $\eta_p^2$ = .003). Further, defiance of advice was punished more when behavior was selfish than prosocial ($b_{\text{selfish behavior}}$ = 3.894; $p$ < .001, 95% CI = [3.622, 4.168], $\eta_p^2$ = .150). Lastly, the interaction between compliance (vs. defiance) and selfish (vs. prosocial) behavior was significant ($b_{\text{comply with advice × selfish behavior}}$ = -1.278; $p$ < .001, 95% CI = [-1.664, -0.892], $\eta_p^2$ = .009), indicating a smaller gap between punishment of prosocial decision-makers who complied with (vs. defied) advice than selfish ones. Finally, for selfish behavior, compliant decision-makers were punished less than defiant decision-makers ($b_{\text{comply with advice}}$ = -1.778; $p < .001$, 95% CI = [-2.051, -1.505], $\eta_p^2$ = .035)[6]. Overall, results indicate that the pattern of punishment for selfish behavior goes above and beyond a simple penalty (reward) of defiance of (compliance with) advice, see Figure 3.

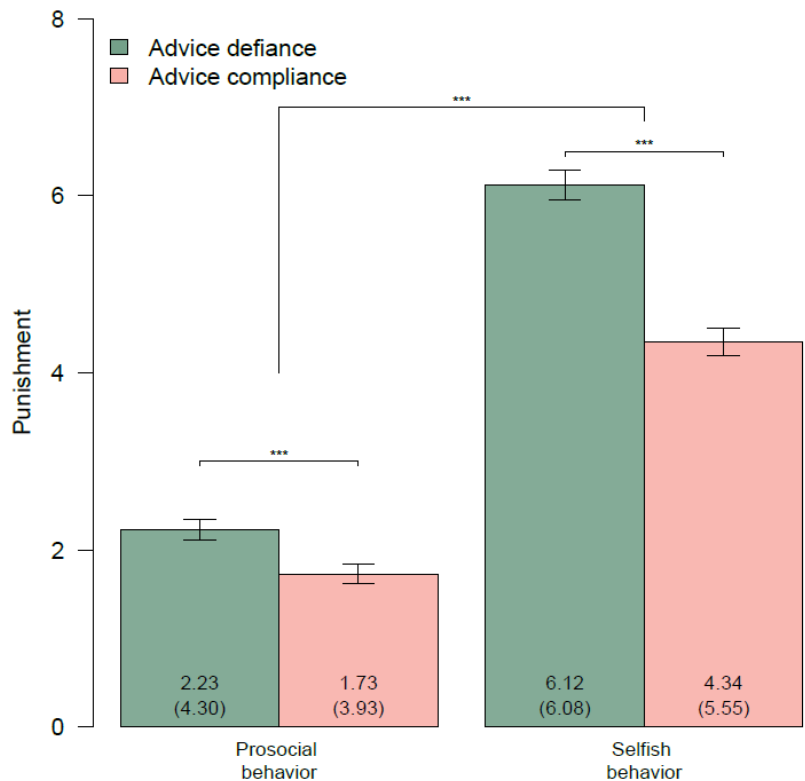

**Figure 3.** Mean points spent to punish decision-makers based on advice defiance vs. compliance and behavior (prosocial or selfish). The comparison between the midpoint of prosocial vs. selfish behavior indicates the significant interaction effect. Mean (*SD*) appears at the bottom of each bar; ***$p < .001$; **$p < .01$, *$p < .05$, +$p < .1$, ns: $p > .1$.

---

[6]To test this gap we ran the same regression, replacing the 'selfish behavior' variable with a 'prosocial behavior' variable. The beta reported is the coefficient of the 'compliance with advice' variable.

Punishment following AI advice

**Perception of responsibility**. On average, responsibility was attributed more to decision-makers than to the advisors ($M$ = 26.86, $SD$ = 27.24), with responsibility rated significantly lower than 50 (the scale midpoint, indicating equal responsibility between the decision-maker and advisor), $b$ = -23.138, $p$ < .001, 95% CI = [-24.698, -21.577][7].

Next, we examined whether perceptions of responsibility varied between human and AI advice based on whether decision-makers complied with or defied the advice. Results reveal that when decision-makers complied with advice, responsibility shifted more away from the decision-maker and towards the advisor when the advisor was human, compared to AI. This pattern of advice source-dependent responsibility shift did not occur when decision-makers defied advice.

Specifically, focusing on prosocial behavior, when decision-makers complied with (prosocial) advice, responsibility scores were higher when advice was written by a human ($M$ = 37.50, $SD$ = 24.37) than an AI ($M$ = 34.95, $SD$ = 26.38, $b_{\text{AI advice}}$ = -2.554; $p$ = .008, 95% CI = [-4.465, -.643], $\eta_p^2$ = .004), indicating responsibility shifted towards the human advisor more than the AI advisor. Further, responsibility scores were higher when decision-makers complied with human advice compared to when they defied it ($M$ = 17.99, $SD$ = 25.08, $b_{\text{defiance of advice}}$ = -19.508; $p$ < .001, 95% CI = [-21.419, -17.597], $\eta_p^2$ = .174), and the interaction between compliance (vs. defiance) and selfish (vs. prosocial) behavior was significant as well ($b_{\text{defiance × AI advice}}$ = 2.949; $p$ = .032, 95% CI = [.246, 5.652], $\eta_p^2$ = .002). When decision-makers defied the (selfish) advice, there was no difference between human and AI advice ($b_{\text{AI advice}}$ = .395; $p$ = .685, 95% CI = [-1.516, 2.306], $\eta_p^2$ < .001)[8], see Figure 4 and Table 2 (model 1).

---

[7]To test the difference, we regress the ['responsibility' score minus 50] on a model which includes only the intercept as an explanatory variable, together with a random intercept for each participant.

Similarly, focusing on selfish behavior, when decision-makers complied with (selfish) advice, responsibility scores were higher when advice was written by a human ($M$ = 37.79, $SD$ = 25.07) than an AI ($M$ = 35.74, $SD$ = 27.15, $b_{\text{AI advice}}$ = -2.046; $p$ = .036, 95% CI = [-3.959, -.133], $\eta_p^2$ = .002), indicating responsibility shifted towards the human advisor more than the AI advisor. Further, responsibility scores were higher when decision-makers complied with human advice compared to defied it ($M$ = 15.94, $SD$ = 23.50, $b_{\text{defiance of advice}}$ = -21.848; $p$ < .001, 95% CI = [-23.761, -19.935], $\eta_p^2$ = .209); the interaction between compliance (vs. defiance) and selfish (vs. prosocial) behavior did not reach significance ($b_{\text{defiance × AI advice}}$ = 2.699; $p$ = .051, 95% CI = [-.005, 5.405], $\eta_p^2$ = .002). When decision-makers defied the (prosocial) advice, there was no difference between human and AI advice ($b_{\text{AI advice}}$ = .654; $p$ = .503, 95% CI = [-1.259, 2.567], $\eta_p^2$ < .001), see Figure 4 and Table 2 (model 2)[9].

---

[8]To test this gap we ran the same regression (Table 2, model 1), replacing the 'defiance advice' variable with a 'compliance with advice' variable. The beta reported is the coefficient of the 'AI advice' variable.

[9]The responsibility results were the same also when pooling the prosocial and selfish behavior together.

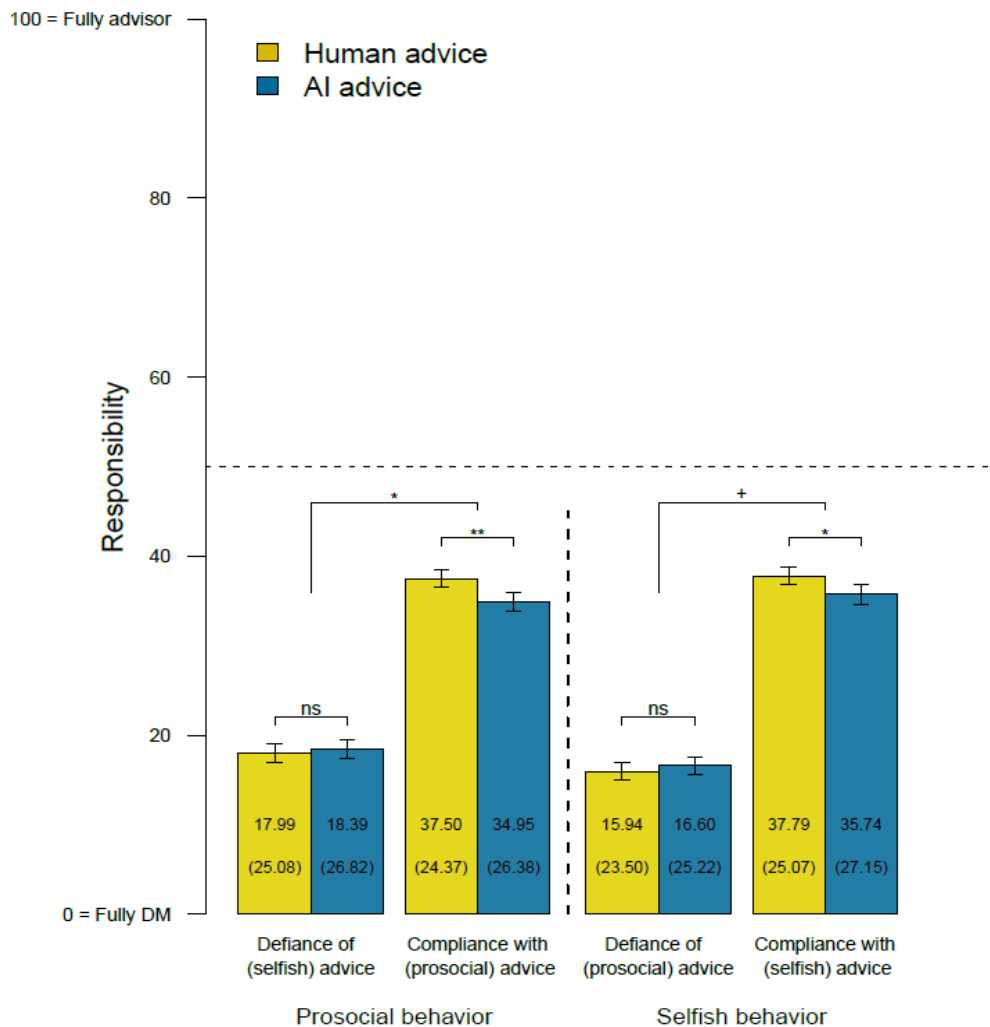

**Figure 4.** Mean reports of perceived attribution of responsibility based on behavior (prosocial vs. selfish), compliance with (vs. defiance of) advice, and advice source (human vs. AI). The comparison between the midpoint of defiance vs. compliance indicates the interaction effect. The means (*SD*) are at the bottom of each bar. The dashed line represents the midpoint (50), in which responsibility is equally shared between the decision-maker (DM) and advisor; ***$p < .001$; **$p < .01$, *$p < .05$, +$p < .1$, ns: $p > .1$.

Punishment following AI advice

| | Dependent Variable: Perception of responsibility | |
|---|---|---|
| | (1) | (2) |
| AI advice | -2.554** (.975) | -2.046*** (.976) |
| Defiance of advice | -19.508*** (.975) | -21.848*** (.976) |
| Defiance of advice × AI advice | 2.949* (1.379) | 2.699+ (1.381) |
| Intercept | 37.50*** | 37.79*** |
| $R^2$ (conditional) | .594 | .595 |
| $R^2$ (marginal) | .111 | .142 |
| $N$ observations | 2532 | 2532 |
| $N$ participants | 633 | 633 |
| Data used for analysis | Prosocial behavior conditions & advice is present | Selfish behavior conditions & advice is present |

**Table 2.** Regression analyses on the perception of responsibility (higher scores = more responsibility of the advisor and less of the decision-maker). $+p < .10$, $*p < .05$, $**p < .01$, $***p < .001$.

***Perception of responsibility among punishers and non-punishers.*** Our results revealed a

discrepancy between the *behavioral punishment* results and the *self-reported perception*

*of responsibility* results. Whereas participants report perceiving a decision-maker as

less responsible (and an advisor as more responsible) for following human (vs. AI)

advice, they do not punish a decision-maker less for following a human (vs. AI) advice. To examine this perception-behavior discrepancy further, we assessed whether the pattern of results regarding responsibility perception differs among participants who engaged in punishment and those who did not.

First, we split our sample into two: non-punishers ($N$ = 155) – participants who spent zero points to punish decision-makers (across all ten punishment decisions), and punishers ($N$ = 478) – participants who spent at least 1 point to punish decision-makers. Then, we examine whether the pattern of attribution of responsibility differs across groups (punishers vs. non-punishers). We extended Model 1 in Table 2 (pooling selfish and prosocial behavior together) by allowing all variables–defiance of advice, AI advice, and the interaction between them–to interact with a dummy variable representing group (punishers vs. non-punishers). This approach allows each group to show distinct patterns of perceptions of responsibility.

We then tested whether a restricted model, where the effect of advice source (human vs. AI) on perceptions of responsibility is the same across groups, fits our data as well as the unrestricted model. A likelihood ratio test failed to reject the null hypothesis, suggesting that the restricted model fits as well as the unrestricted model, $\chi^2(2)$ = .745; $p$ = .688. Thus, our results indicate that punishers and non-punishers show similar patterns of responsibility perception when it comes to human vs. AI advice sources[10].

These results suggest that the perception-behavior discrepancy is not driven by a specific type of participant. If, for instance, (i) punishers would exhibit a responsibility pattern similar to their punishment pattern (i.e., equal levels of decision-makers' responsibility when they follow AI and human advice), and (ii) non-punishers would exhibit a different responsibility pattern (i.e., higher perceptions of decision-makers' responsibility when they follow AI vs. human advice), we could have concluded that the overall pattern of responsibility perception in our data, and the perception-behavior discrepancy is driven by participants who do not necessarily take the task seriously or care about punishing others. However, given that the pattern of responsibility perception is similar among those who punish and those who do not, it seems like other motivations drive the perception-behavior gap. In the discussion section, we introduce the ideas of cost-benefit analysis, social desirability, and how the questions are asked as potential reasons for the perception-behavior gap.

## Discussion

Recent incidents from daily life show that people misuse AI advice for personal gain, and that punishment of such behavior varies dramatically (Merken, 2025; Cole, 2025). As such, the current study takes an experimental approach and examines how much people are willing to punish selfish behavior that occurs after AI advice. It further compares the level of punishment to that of selfish behavior that occurs after human or no advice, and to the punishment of prosocial behavior (after no advice, human advice, or AI advice). Consistent with our first hypothesis, we find that prosocial behavior is punished rather little, whereas selfish behavior is punished much more. This finding confirms the vast literature on the punishment of antisocial behavior, demonstrating yet again that punishment is a deterrence tool used to enforce and uphold prosocial norms (Balliet et al., 2011; Balliet & Van Lange, 2013).

---

[10] In the SOM we report an additional exploratory analysis, testing whether the pattern of perceptions of responsibility varies across other types of punishment behavior. Results reveal no such difference, see SOM.

Our results further reveal that the type of advice individuals receive shapes their punishment, especially when they engage in selfish behavior. Selfish behavior was punished more when it occurred after prosocial advice and less when it occurred after selfish advice. Based on attribution theory (Heider, 2013; Kelley, 1973) and related work (Monroe et al., 2015; Pavey & Sparks, 2009), any behavior is considered more autonomous when it is independent and does not occur after persuasion. In line with this theory, selfish behavior should be seen as more autonomous when it occurs after prosocial advice and less autonomous when it occurs after selfish advice. Indeed, in our study, participants perceived selfish decision-makers as less responsible (and their advisors as more responsible) when they read selfish (vs. prosocial) advice. Consequently, participants also punished these decision-makers less.

Examining the effect of the source of advice, we found that punishment levels did not differ when an AI or a human provided decision-makers with advice. Recent work examined the attribution of blame and responsibility for humans and machines by mostly employing hypothetical scenarios (Arnestad et al., 2024; Aschauer et al., 2023;

Wilson et al., 2022). Much of this work revealed that machines are less blameworthy than humans for moral transgressions. Our work, however, is the first to explore *behavioral, costly* punishment (Fehr & Gächter, 2004), revealing a different story. Namely, when it comes to *costly* punishment, individuals are not willing to forgo their pay to punish those who follow selfish AI advice more than those who follow selfish human advice. We thus add to the emerging literature that shows that, in some situations, machines and humans are blamed, held accountable, and relied upon to the same extent when they engage in wrongdoings (Carrasco-Farre, 2024; Leib et al., 2024; Lima et al., 2021; Malle et al., 2025).

Whereas evaluators punish behavior similarly when it occurs after human or AI advice, perceived responsibility differs between the two advice sources. The exploratory perception results show that evaluators attributed more responsibility to human advisors than AI advisors when decision-makers complied with the advice. The pattern of perceived responsibility aligns with legal perspectives on coercion, where the degree of influence over a decision affects culpability (Dressler, 2001; Feinberg, 1989). From a legal standpoint, responsibility for an action is often shared or even shifted toward the source of persuasion when an individual is seen as having acted under undue influence. AI, lacking intent and agency, may be perceived as less capable of exerting true coercion, leading evaluators to place greater responsibility on human advisors. This suggests that while AI-generated advice can alter moral decisions, it may not be judged as morally coercive in the same way human persuasion is judged (Constantinescu et al., 2022).

The pattern of perceived responsibility differs from the pattern of punishment, resulting in a perception-behavior gap. Decision-makers who follow human (vs. AI) advice are seen as less responsible, but both are equally punished. Note that the responsibility results are exploratory and require further validation and replication. Nevertheless, we consider several potential reasons for the obtained perception-behavior gap.

Baumert and colleagues (2013) propose a moral courage model, describing the stages required for a person to intervene against a norm violation. First, an individual should be able to detect and interpret a behavior as a norm violation. In our setting, participants are confronted with the (selfish) behavior and indeed interpret it as selfish (see footnote 2 and SOM for details). Then, individuals should feel responsible and capable of intervening. Our experiment puts participants in the role of a punisher by explicitly asking them to choose how much to punish, thereby highlighting their responsibility. Further, by providing participants with additional points (later converted to money), the experiment makes them capable of punishing decision-makers.

The final stage of the moral courage model is a cost-benefit analysis, where individuals conduct a private analysis weighing whether the benefits of intervening are superior to the costs of staying silent. A similar cost-benefit analysis process was proposed by Molho and colleagues (2019) who examined punishment strategies in daily life. If, indeed, our experimental set-up clears all the previous stages, the final cost-benefit analysis stage may account for the perception-behavior gap. It might be that evaluators *say* they perceive decision-makers as more responsible when they follow AI (vs. human) advice, but are simply unwilling to 'pay the price' of higher levels of punishment. Namely, they might not be willing to forgo more of their own financial endowment to punish decision-makers.

Further, it might be that participants' self-report of perceived responsibility is not accurate or sufficiently sensitive. Namely, evaluators might not report their actual perception of responsibility but report values due to social desirability concerns, a common challenge in self-report measures (van de Mortel, 2008). Alternatively, evaluators might report their perceptions accurately, but the gaps in perception might not be large enough to translate to a corresponding gap in the costly punishment decision. Overall, the perception-behavior gap obtained here highlights the importance of supplementing research that examines self-reports with studies employing incentivized behavioral measures. Combining the two approaches can facilitate a more comprehensive and nuanced understanding of judgment and punishment in hybrid human-AI settings.

Finally, the perception-behavior gap might stem from the way the questions were presented to participants. When asking about perceptions of responsibility, the question included a scale on which evaluators had to split responsibility between the decision-maker and the advisor. This presentation might have encouraged evaluators to view responsibility as a fixed pie and consider both the decision-maker and the advisor when replying. When asking about punishment decisions, however, evaluators were only asked to punish the decision-maker. As such, participants might have paid less attention to the advice source when making punishment decisions. Indeed, in a follow-up survey, 34% of participants who correctly predicted the punishment results indicated that the focus should be on the decision-makers and not advisors when punishing (see SOM).

The decision to not introduce costly punishment to the advisor allowed us to compare punishment of behavior after (human and AI) advice to a control condition where advice was not present. Further, implementing costly punishments for AI advisors is not trivial. Previous work shows that different payoff implementations shape social preferences toward machines (von Schenk et al., 2023), indicating that the punishment of AI advisors might be contingent on how AI advisors are presented and how their payoffs are implemented in the experimental design. As such, examining how variations in AI presentation and payoff implementation shape costly punishment decisions presents an important next step for future research.

### Contribution and Novelty

Our work contributes to the field on multiple levels, from specific ones related to the studied topic, to broader ones related to the field as a whole. At the localized level, we advance the study and knowledge of responses to ethical violations in human-AI settings. Previous research has largely relied on hypothetical scenarios and self-reported measures to examine these reactions (e.g., Awad et al., 2020; Bigman et al., 2019; Lima et al., 2021; Malle et al., 2025). Utilizing these methods helps to ground participants in realistic settings, increasing ecological validity. At the same time, such measures carry limitations such as social desirability bias (Van de Mortel, 2008), and divergent results between self-reported and actual punishment behavior (Baumert et al., 2013). To address these gaps, our work adopts a behavioral economic approach, employing an incentivized, behavioral punishment measure (Fehr & Fischbacher, 2004; Pillutla & Murnighan, 1996). Doing so allowed us to capture punishment that is consequential (both to the punisher and the one being punished), and mitigate social desirability concerns. At the same time, making punishment costly may overestimate levels of punishment compared to non-costly punishment, as people might be more willing to punish when they do not need to directly confront those punished (Molho et al., 2020). Conversely, costly punishment might underestimate levels of punishment because it requires individuals to forgo part of their own pay to punish. Given that different methodologies have distinct strengths and weaknesses, triangulating methodological approaches is key for a comprehensive understanding of punishment in human-AI settings. Future work should build on others' and our work and supplement current results by employing alternative methods for measuring punishment, such as diary studies (Molho et al., 2020), and analyses of field data (Hebl et al., 2007; Jacobsen, 2020).

Beyond its localized contribution, our work also advances key theories in social psychology. First, the results obtained here contribute to the classic attribution theory (Heider, 2013; Kelley, 1973), expanding it to settings with novel technology. We find that not only does *human* persuasion shape perceptions of responsibility (and punishment) but that AI, as a persuasive tool, has a similar effect. AI advisors' ability to increase (when the advice promotes prosociality) and decrease (when the advice promotes selfishness) punishment for wrongdoings similarly to human advisors marks a first step in extending the classic attribution theory into the new age of hybrid human-AI interactions.

Second, the findings contribute to theories of punishment and norm enforcement by demonstrating that people punish selfish behavior according to the advice content and not its source. The fact that AI and human advisors result in similar levels of punishment suggests that punishment, as a norm enforcement mechanism, extends beyond interpersonal human-human dynamics (Balliet et al., 2011; Balliet & Van Lange, 2013). As such, our results support theories of norm compliance that emphasize the general context and expectations of certain behaviors, over the specific agents involved in norm violations (Bicchieri, 2006; Cushman, 2015).

More broadly, this work contributes to the growing field of psychology of AI ethics. Prior research has explored AI's impact on altruism and reciprocity (Köbis et al., 2024), delegation of moral decisions (Köbis et al. 2024), dishonesty (Leib et al., 2024), and accusations of lying (von Schenk et al., 2024). We extend this work by adding insights into how AI-advisors shape punishment, a key mechanism in enforcing social norms. Our data suggest that if individuals recognize that moral transgressions are punished also when guided by AI, they may be less inclined to use AI as a justification for selfish behavior and opt for prosocial behavior instead. Future work should systematically examine how AI, in various roles (e.g., advisor, role model, partner, see Köbis et a., 2021), affects moral outcomes and test interventions that reduce AI-driven moral transgressions. By doing so, the field can ensure that individuals benefit from the potential of AI, while upholding high moral standards.

At the broadest level, this research contributes to the study of human-AI interactions by adding to the ongoing debate on whether AI should be viewed as a moral agent or merely as a tool. Whereas the philosophical debate on the issue is split (Maruyama, 2022; Heinrichs, 2022; Constantinescu et al., 2022), our findings suggest that, in practice, people attribute some degree of moral agency to AI. Namely, we show that individuals assign some blame to AI advisors rather than holding only the human decision-maker responsible for wrongdoing. Further, they punish selfishness similarly, regardless of the advice source. Given that AI is increasingly embedded in human decision-making, perceptions of AI's moral agency will shape how AI systems are used, designed, and regulated.

Lastly, our findings contribute to the growing body of interdisciplinary work linking psychology, economics, machine behavior, philosophy, and computer science to better understand AI's impact on humans. In our work, we adopt a machine behavior approach (Rahwan et al., 2019), building on theories from social psychology (Heider, 2013; Kelley, 1973) and employing methods from behavioral economics (Fehr & Fischbacher, 2004). As AI becomes an integral part of human decision-making, interdisciplinary research becomes evermore important for addressing complex questions around the integration of AI and its impact on humans. Interdisciplinary work is particularly relevant when studying human-computer interactions because each field contributes unique perspectives and methodologies: philosophy provides ethical frameworks, computer science offers technical insights and capabilities, psychology contributes insights on humans' underlying processes, and economics helps by systematically modeling humans' behavior and reactions to incentives. Overall, borrowing from different fields and combining their strengths will lead to a more comprehensive understanding of human-computer interactions.

### Future Directions and Implications

Our study examined punishment preferences in a representative US sample, allowing for a broader conclusion beyond specific demographic groups (Dong et al., 2024; Henrich, 2023). However, the extent to which these findings generalize beyond the US remains an open question. While punishment of selfish behavior is universal, its severity varies across cultures, with larger and more altruistic societies enforcing prosocial norms more strongly (Henrich et al., 2006). Attitudes toward AI also differ across cultures (Dong et al., 2024; Ma et al., 2024; Yam et al., 2023; Globig et a., 2024), potentially shaping how decisions following AI-advice are judged and punished. For instance, collectivist societies, that value social harmony, may view AI advice as a socially endorsed tool, leading to more lenient punishment. On the other hand, societies more skeptical of AI may impose stricter punishment. Future research should examine punishment in hybrid human-AI settings across diverse cultures to test the universality of our findings and uncover potential cultural moderators.

From a practical standpoint, the findings obtained here provide valuable insights for organizations and policymakers. Organizations that use AI systems in advisory roles should expect a strong public reaction to selfish outcomes caused by employees following AI advice. Public scrutiny might be as strong as it would be for employees following similar human advice. Given the increased reliance on AI advisors in the workplace, organizations should take extra care to ensure that the AI systems they employ provide ethical advice that serves the public.

Furthermore, legal systems must better define responsibility and associated consequences (i.e., punishment) in hybrid human-AI interactions. This is especially true as current legal frameworks struggle to do so (Smith et al., 2024). As a case in point, a court recently held an airline liable for its chatbot providing incorrect and costly advice to a customer despite the airline denying responsibility for the chatbot's bad advice (Zittrain, 2024). This case demonstrates the new challenges that legal frameworks face in assigning responsibility and punishment in hybrid settings. Our findings show that lay people punish behavior similarly, regardless of whether it stems from AI or human advice. Policymakers should establish regulations that address the question of shared responsibility in hybrid settings, ensuring legal frameworks keep evolving together with technological advancements and take into account people's preferences.

## Conclusions

As AI increasingly influences human decision making, understanding how people punish selfish behavior in hybrid human-AI settings bears immense relevance. Taking a psychological approach to AI ethics, our findings reveal that whereas decision-makers are perceived as more responsible for selfish behavior when they follow AI (vs. human) advice, this does not translate to more punishment. Punishment is driven by the behavior and advice type, but not the advice source. Given the widespread use of AI advisory tools, it is essential to program AI tools to avoid corrupt advice and craft environments where following bad advice is disincentivized. Doing so will assist in upholding positive social norms in hybrid human-AI settings.